# An image sensor based on single-pulse photoacoustic electromagnetic detection (SPEED): a simulation study.

**Authors:** J. Aguirre[1*].

**Affiliation:** Department of Electronics and Communications Technology, Universidad Autónoma de Madrid, Madrid, Spain.

**\*Email:** juan.aguirre@uam.es

## Abstract

Image sensors are the backbone of many imaging technologies of great importance to modern sciences, being particularly relevant in biomedicine. An ideal image sensor should be usable through all the electromagnetic spectrum (large bandwidth) since for every region of the electromagnetic spectrum different tissue parameters can be retrieved, it should be fast (millions of frames per second) to fulfil the needs of many microscopy applications, and it should be cheap, in order to ensure the sustainability of the healthcare system. However, current image sensor technologies have fundamental limitations in terms of bandwidth, imaging rate or price. In here, we briefly sketch the principles of an alternative image sensor concept termed Single-pulse Photoacoustic Electromagnetic Detection (SPEED). SPEED leverages the principles of optoacoustic (photoacoustic) tomography to overcome several of the hard limitations of today's image sensors. Specifically, SPEED sensors can operate with a massive portion of the electromagnetic spectrum at high frame rate (tens of millions of frames per second) and low cost. Using simulations, we demonstrate the feasibility of the SPEED methodology and we discuss the step towards its implementation.

## Introduction:

Imaging systems are of great importance to modern science and society in general, having countless applications across practically all the scientific disciplines and being particularly important in biomedicine. Their backbone is the image sensor. Such sensor detects and conveys information from electromagnetic radiation, which has previously interacted with a sample, to obtain static images or videos of that sample.
The performance of an imaging system can be largely characterized by 1) the bandwidth of the image sensor, defined as the region of the electromagnetic spectrum where the detector has "good" sensitivity and 2) the frame rate (imaging speed) of the image sensor, usually defined in frames per second (fps). Image sensors that use large portions of the electromagnetic spectrum, that reach high speeds ($>10^6$ fps), are in strong demand by the scientific community in general, and the biomedical community in particular.

Commonly available image sensors can be classified into photoelectric image sensors and thermal image sensors. The active elements (pixels) of photoelectric image sensors are semiconductors and rely on photoelectric effects produced by the incoming radiation [1]. The development of silicon-based charge-coupled devices (CCD) and complementary metal-oxide-semiconductor (CMOS) pixelated sensors has brought the benefits of cheap, high-performance, imaging sensors that have become ubiquitous in technologies ranging from mobile phones to professional digital single-lens reflex (SLR) cameras. Silicon-based CCDs and CMOS exploit the high-throughput fabrication techniques that are widely used in the semiconductor industry. However, due to the physics of the bandgaps of semiconductors, only photons in the visible induce an electrical signal, limiting the spectral response range. Moreover, the complexity of the electronic readout process prevents from achieving frame rates beyond $10^6$ fps. More-over at $10^6$ fps only a few frames can be obtained.

In thermal image sensors, radiation is absorbed by pixels, which increases their temperature, triggering a change in an electronically measurable physical property of the pixel e.g. its resistance. These image sensors are generally insensitive to the wavelength of the incident radiation, but they are very slow (typically milliseconds, $10^3$ fps), due to the time it takes for the pixel to heat or cool. In addition, thermal detectors generally have to be cooled to cryogenic regions when operating in the infrared and terahertz range to avoid noise due to spontaneous background emission [1]. Overall, thermal image sensors have not matured into cheap multipixel devices with good timing response like their silicon-based counterparts.

Alternatively, the cost limitations of multipixel thermal image sensors can be overcome by implementing single-pixel imaging strategies. The simplest strategy consists on raster scanning the illumination source or a single-pixel. More advanced methods use spatial light modulators to illuminate the scene with spatially resolved patterns while measuring the correlated intensities using a single-pixel detector, offering improvements in information compression [2]. Nevertheless, such approaches do not overcome the fundamental drawbacks in terms of imaging speed and spectral abilities described above. Moreover, cheap commercially available spatial light modulators are only available for the visible region. Furthermore, the need of projecting patterns also adds significant limitations to the imaging speed. To the best of my knowledge the fastest method could only achieve 72fps in the visible region [3].

In here we propose a new image sensor paradigm termed Single-shot Photomechanical Electromagnetic Detection (SPEED). This scheme leverages the transduction of the illumination radiation onto photoacoustic waves. By making use of the principles of optoacoustic tomography, such an image sensor offers substantial advantages in terms of bandwidth, background noise suppression and electronic read-out complexity. More specifically, SPEED image sensors can produce images using the whole electromagnetic spectrum at much higher imaging rate than classical thermal sensor, constant background radiation sources do not affect the sensitivity of SPEED image sensors, the complexity of the readout electronics can be drastically reduced in SPEED sensors making their fabrication very cost efficient.

In this paper we briefly describe the theory and simulations that proof the feasibility of the SPEED image sensor concept. Our simulations indicate that a sensor usable at the whole electromagnetic spectrum, using a single ultrasound detector for electronic read-out at imaging rates of ~$10^6$ fps is feasible.

# Methods:

## Advantages of the SPEED image sensor method.

In this section we explain the foundations of the SPEED concept and its advantages.

The SPEED scheme works by actively illuminating the imaged object using a time-varying electromagnetic source. The radiation reflected, transmitted, scattered or reemitted by the object is directed using a focusing element (objective) onto a radiation-absorbing pixels, where transient heat absorption occurs, producing a photoacoustic (optoacoustic) wave. In this transduction operation, the fluence rate of the electromagnetic radiation at the pixel ($\Phi(r,t)$, $W/cm^2$) is encoded into a "slow-travelling" ultrasound wave ($p(r,t)$, $N/cm^2$). The ultrasound wave is detected by an ultrasound sensor and then digitized. The information related to $\Phi(r,t)$ of interest for image formation can then be calculated from the detected $p(r,t)$ (see Figure 1b and c).
Such an image sensor offers the following substantial advantages:

**a)** SPEED image sensors can produce images using the whole electromagnetic spectrum in similar fashion to classical thermal image sensors. The non-radiative atomic transitions involved in the photoacoustic process usually have a mean life time of around ~1 ps and provide the ultimate speed limit of the SPEED concept (~$10^{12}$ fps). The transient heat deposition process encodes the fluence of the illumination radiation in the pixel into a photoacoustic wave.

**b)** Constant background radiation sources do not affect the sensitivity of SPEED image sensors, since the photoacoustic effect is produced only by the absorption of a time-varying electromagnetic source [4]. This is of special interest for regions of the electromagnetic spectrum in which background radiation is very intense, like THz of infrared.

**c)** The complexity of the readout electronics can be drastically reduced in SPEED sensors making their fabrication very cost efficient. The photoacoustic waves coming from different pixels can be integrated in a set of ultrasound detectors (see Figure 1b) or even a single ultrasound detector (see Figure 1c). By using tomographic principles, the amount of radiation absorbed by the pixels can be obtained. The single detector capabilities of the SPEED image sensors are very cost efficient, since for its fabrication one only needs a coupling media (for example silicon), light absorbers sheets (for example carbon sheets) and an ultrasound transducer.

## Working principles.

In this section we present a particular solution of the optoacoustic wave equation. From this equation the implementations of the SPEED concept can be derived.

Let us assume that a time-varying electromagnetic source illuminates a sample and the transmitted/scattered/reemitted radiation is directed by a focusing element (objective) towards light-absorbing pixels (Figure 1b). An ultrasound detector is placed at position $\vec{r}$, and an acoustic coupling medium "connects" the pixels with the transducer. From general optoacoustic theory, [4] we know that some radiation will be absorbed at each pixel producing a temperature increase, which in turn produces mechanical stress that leads to an optoacoustic wave. This process is described by the following equation

$$\left(\nabla^2 - \frac{1}{v_s^2}\frac{\partial^2}{\partial t^2}\right)p(\vec{r},t) = -\frac{\beta}{C_p}\frac{\partial H(\vec{r},t)}{\partial t} \quad (1)$$

where $p(\vec{r},t)$ is the optoacoustic pressure ($N/cm^2$); $v_s$, the speed of sound; $C_p$, the specific heat capacity at constant pressure; $\beta$, the Grüneisen parameter (dimensionless); and $H$, the heating function ($W/cm^3$), defined as the thermal energy deposited per unit volume and per unit time. Let us assume that the electromagnetic radiation is pulsed and that the heat and stress confinement conditions are met [5]. If we assume that the radiation has a delta time profile, the following equation can be derived from equation (1):

$$p_\delta(\vec{r},t) = \frac{1}{4\pi v_s^2}\frac{\partial}{\partial t}\left[\int d\vec{r}'\frac{p_o(\vec{r}')}{|\vec{r}-\vec{r}'|}\delta\left(t-\frac{|\vec{r}-\vec{r}'|}{v_s}\right)\right] \quad (2)$$

where $p_\delta(\vec{r},t)$ is the photoacoustic wave generated by a pulse with a delta shape in time, and $p_o(\vec{r}')$ is the initial pressure. In addition, $p_o(\vec{r}') = \beta\phi(\vec{r}')\mu_a(\vec{r}')$ where $\phi(\vec{r}')$ is the fluence ($J/cm^2$) and $\mu_a(\vec{r}')$ is the absorption coefficient ($1/cm^2$), which can be assumed to be constant at pixel locations and zero everywhere else. Therefore, if $\vec{r}'$ belongs to a point in space occupied by pixel material $p_o(\vec{r}') = \Omega\phi(\vec{r}')$, where $\Omega=\mu_a\beta$=constant and $p_o = 0$ everywhere else.

Let us assume a SPEED image sensor in which the ultrasound transducers and absorbing pixels are placed in such a way that two or more ultrasound waves can arrive at each ultrasound detector at overlapping time points (Figure 2b). Using tomographic algorithms, the amount of radiation absorbed at the pixels is obtained ($\phi(r)$ can be obtained inverting equation (2)). The fact the surface integral in equation 2 is zero everywhere except where the pixels are, greatly reduce the difficulty of the inverse problem. In fact, only a few detectors are needed in order to obtain $\phi(r)$ simplifying the complexity of the readout electronics of SPEED sensor. The acquisition speed is given by the repetition rate of the laser and the size of the absorbing pixels. Alternatively, by adding ultrasound scatterers (Figure 2c) the number of ultrasound transducers can be reduced to one. In this case, equation 2 does not apply anymore and the inverse problem has to be solved using a model matrix that can be calculated using simulations or experimentally, similarly as is calculated for single element photoacoustic imaging in deep tissue [6] [7]. It is important to understand that when adding the scattering elements, the acquisition time will be given by the time length of the recorded signal (~$10^6$ fps assuming that the maximum distance travelled by the ultrasound is 1mm), provided that it is larger than the repetition rate of the laser.

## Simulations

To demonstrate that the SPEED sensor arrangement consisting on a few ultrasound detectors can lead to high fidelity images, we simulated the optoacoustic signal generated when imaging a scenario for different configurations of pixels and detectors.

More specifically, we simulated a SPEED image sensor consisting on 300 x 300 absorbing pixels with an optical absorption coefficient of 1 (a.u), each of them being a sphere of 10 micrometre diameter and being adjacent to each other (see Figure 2a,b,c). We assumed that the ultrasound coupling media is water and that the ultrasound detectors are situated 1 mm above the light absorbing pixels. We simulated three scenarios:

Scenario 1: the number of ultrasound detectors is half the number of absorbing pixels (Figure 2a).
Scenario 2: the number of ultrasound detectors is 1/6 the number of absorbing pixels (Figure 2b).
Scenario 3: the number of ultrasound detector is 1/20 the number of absorbing pixels (Figure 2c).

We reconstructed the images from the optoacoustic data, inverting equation 2 by using the model based reconstruction algorithm presented in [8]. The inverse problem was solve using the least squares method (lsqr).

To demonstrate that the arrangement consisting on the combination of scatterers and a single ultrasound detector can lead to high fidelity images, we simulated the optoacoustic signal generated from 7 light absorbing pixels forming a linear array. Each of the pixels received a random amount of radiation (see Figure 2d and 2e). We assume point like pixels separated 0.6 mm from each other. A point like detector was placed 2.1 mm above the line array at the same x coordinate as the central radiation absorbing pixel (Figure 2b). We simulated two scenarios:

Scenario 4: the ultrasound coupling medium is water and it is homogenous.
Scenario 5: the ultrasound coupling medium is water containing an air disc that acts as a scatterer breaking the symmetry of the system.

The model matrix was calculated empirically as it is done in similar problems devoted to single detector optoacoustic imaging of deep tissue [6] [7]. We performed all the simulations and built the matrix using the matlab toolbox K-wave [9] . The inverse problem was solved using lsqr.

# Results

In Figure 1 we show the schematics of the working principle of a standard image sensor and the SPEED concept. The working principle of a standard image sensor is shown in Figure 1a. The working principle of a SPEED sensor using a homogenous ultrasound coupling media and several ultrasound transducers is shown in Figure 1b. Last but not least, the working principle of a SPEED sensor using a scatterers in the coupling media and several ultrasound transducers is shown in Figure 1c.

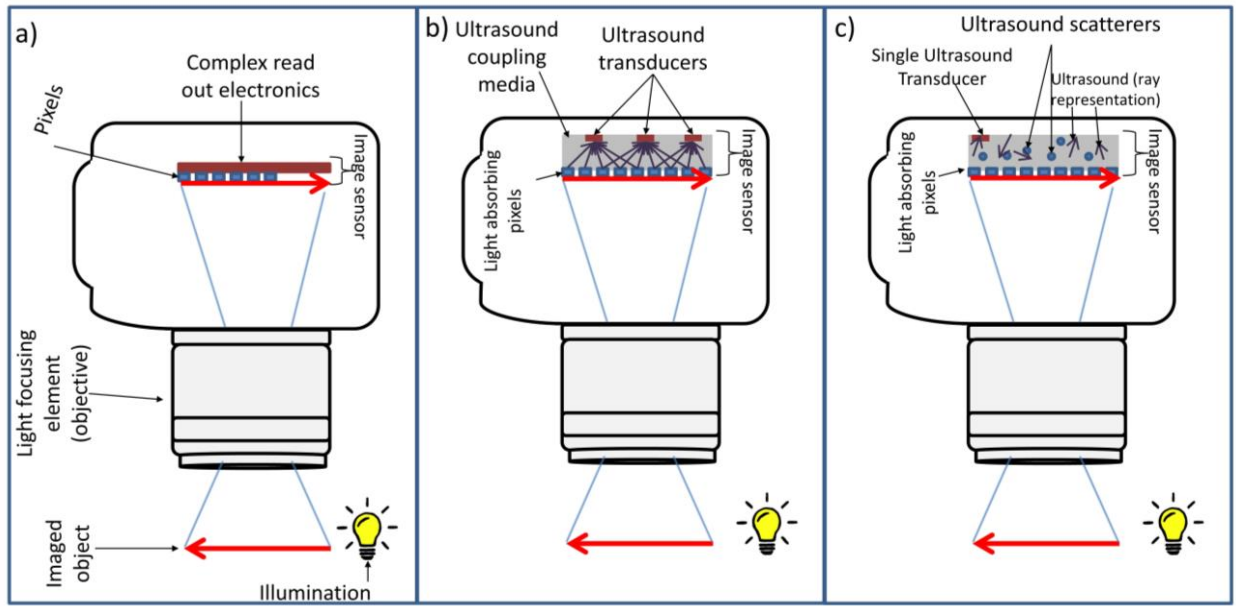

**Figure 1**: **Schematic of the working principle of the SPEED sensor concept**. (**a**) A classical image sensor inside a camera. The illumination radiation is directed onto the imaged object. The scattered/transmitted/reemitted radiation is focused by an objective, which projects an image of the object onto the image sensor. The photons induce a change in the electrical properties of the pixels. Such a change is "read out" by a complex electronic system. (**b**) A SPEED image sensor inside a camera. The scattered/transmitted/reemitted radiation is generated as a time-varying radiation signal arriving at the pixels, where transient heat deposition occurs. In each pixel a photoacoustic wave is generated. The information of the time profile of the scattered/transmitted/reemitted radiation is encoded in photoacoustic waves which travel through an acoustic coupling medium. Several ultrasound transducers detect the ultrasound waves. Two or more ultrasound waves can arrive at the detector at overlapping time points. Using tomographic algorithms, the amount of radiation absorbed at the pixels can be obtained (**c**). Another SPEED sensor implementation using a single ultrasound detector. In each pixel, a photoacoustic wave is generated. One ultrasound transducer detects the ultrasound waves which undergo multiple scattering before reaching the detector. Like in b) using tomographic algorithms, the amount of radiation absorbed at the pixels can be obtained.

In Figure 2 we show the results of the simulations that proof the feasibility of the SPEED concept.

The results of the simulations for the scenarios in which the number of pixels is greater than one are shown in Fig.2a,b and c. The figure includes a scheme of the distribution of pixels and ultrasound detectors, the raw sinograms and the reconstructed images for scenarios 1,2 and 3 are shown in Figure 2a,b and c respectively. When the number of radiation absorbing pixels is two times the number of ultrasound detectors, the image is correctly reconstructed (scenario 1, Fig. 2a). This is still the case when the number of radiation absorbing pixels is 6 times the number of ultrasound detectors (scenario 2, Fig. 2b). However, when then number radiation absorbing pixels is 20 times the number of ultrasound detectors the image cannot be reconstructed (scenario 3, Fig. 3b).

The result of the simulations in which there is only one ultrasound detector are displayed in Fig.2d,e.
As expected, for the case in which a scatterer breaks the symmetry in the acoustic coupling media the retrieved radiation intensity values were correct. For the homogenous ultrasound coupling media case, the radiation intensity values were not retrieved correctly for most the pixels.

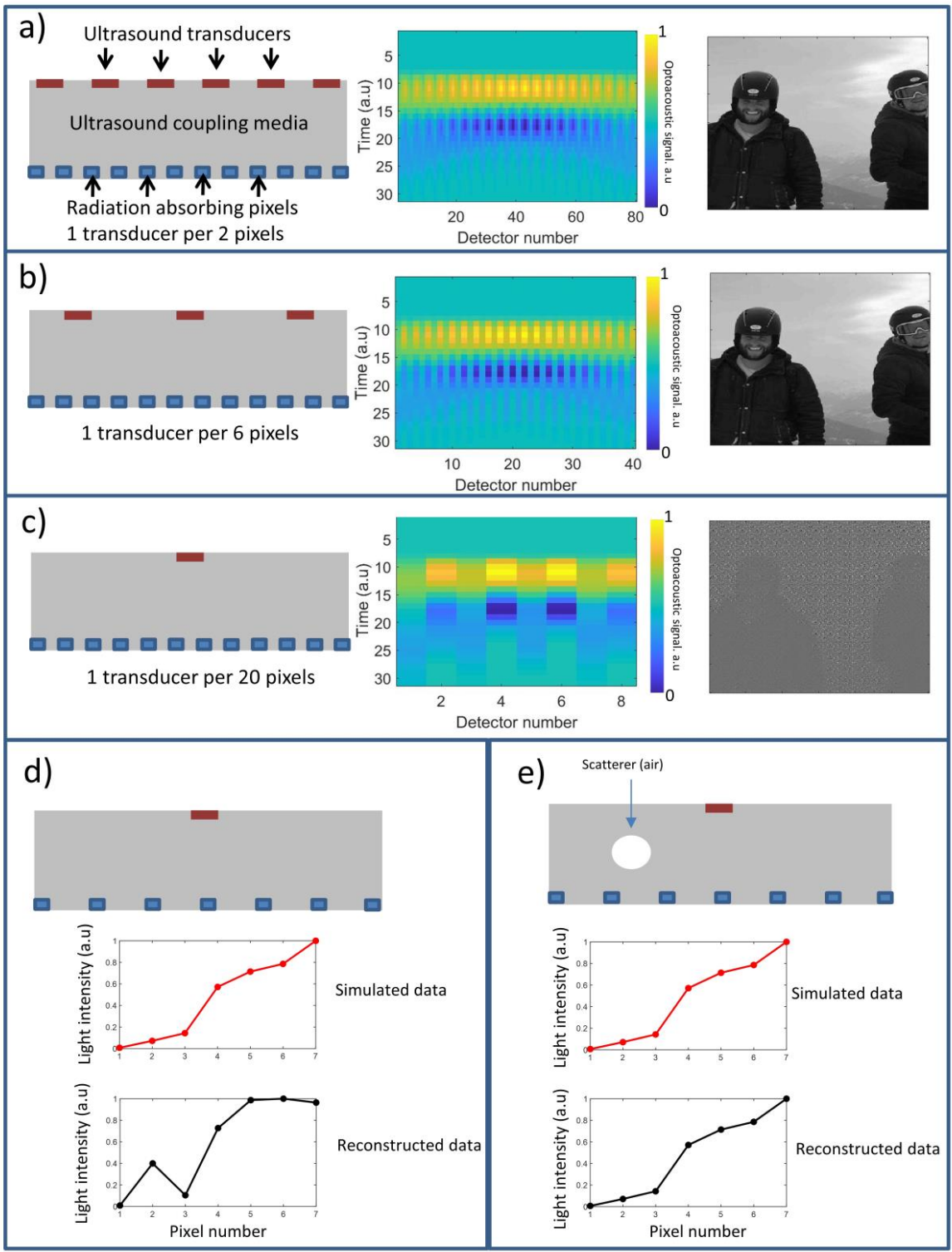

**Fig 2**. **Simulation demonstrating that the number of ultrasonic transducers can be several times smaller than the number of light-absorbing pixels or a single one.** (**a**) *left:* configuration of ultrasound transducers and pixels (2 pixels per transducer)**.** *Middle:* optoacoustic signal recorded by each transducer (sinogram)**.** *Right:* reconstructed image with no quality loss. (**b**) *Left:* configuration of ultrasound transducers and pixels (6 pixels per transducer). *Middle:* optoacoustic signal recorded by each transducer (sinogram). *Right:* reconstructed image with no quality loss. (**c**) *Left:* configuration of ultrasound transducers and pixels (20 pixels per transducer). *Middle:* optoacoustic signal recorded by each transducer (sinogram). *Right:* reconstructed image showing quality loss. (**d**) *Top:* configuration using a single ultrasound detector and seven light absorbing pixels. *Middle:* light intensity values in each pixel use for the forward simulation. *Bottom:* Reconstructed values. The inverse problem algorithm (least squares method) was not able to retrieve the correct light intensity values in each pixel. (**d**) *Top:* configuration using a single ultrasound detector and seven light absorbing pixels and a scattering region. *Middle:* light intensity values in each pixel use for the forward simulation. *Bottom:* Reconstructed values. The inverse problem algorithm (least squares method) the scatterers break the symmetries of the signals and allows for correct retrieval of the pixel values.

# Discussion.

We have proposed a new image sensor model (SPEED) and demonstrated using simulations its basic principles. SPEED uniquely combines usability in the whole electromagnetic spectrum, high frame rate, and low cost. The method has the potential to overcome the limitations of current image sensors for active illumination applications.

The large bandwidth is due to the fact that SPEED is based on the absorption of time varying radiation by pixels leading to heat deposition which in turn creates an optoacoustic wave. Heat deposition by radiation occur through the whole electromagnetic spectrum for different kind of materials. In fact, the detection mechanism of SPEED resembles very much the one of thermal sensors. In such sensors, radiation is absorbed by pixels, which increases their temperature, triggering a change in an electronically measurable physical property of the pixel e.g. its resistance. However, the frame rate limitation of a single pixel SPEED sensor is orders of magnitude higher since it is based on transient heat deposition in the absorbing pixel while thermal image sensors need to achieve thermal equilibrium to retrieve the parameter of interest. Therefore, a SPEED sensor does not have anything to do with a classical thermal image sensor.

The low cost is due to the fact that by taking advantage of the principles of photoacoustic tomography the electronic readout system consists simply in one or few ultrasound transducers are needed. In fact, to build a SPEED sensor one only needs an ultrasound transducer, acoustic coupling medium (for example water) and radiation absorbers (for example carbon).

The advantages over standard sensors in terms of frame rate stands also from the low complexity of the read out electronics. Standard image sensors based on CCDs o CMUTs need complex electronic readout schemes to retrieve the electronic signal generated at each pixel. As a result, their image rate limit is only around 1Mfps. At those imaging rates, only a few frames can be retrieved. However, for the SPEED imaging case the complexity of the readout electronics is reduced to the complexity of one or few ultrasound transducers. For the simulations presented in this work the maximum frame rate is given by the characteristic length of the (case of Figure 2b) or the time length of the photoacoustic signal (case of Figure 2c) being around 50Mfps and 1Mfps respectively.

Once the general principles in terms of spectral range, frame rate and cost of the SPEED concept are proofed the next question is: what is the sensitivity of a SPEED sensor for each spectral window? Other parameters like pixel size, dynamic range, minimum resolved energy play a major role for each imaging application. There is great versatility about the possible designs of the SPEED concept in terms of geometries and possible radiation absorption materials to be used as pixels and ultrasound detection technology, therefore, we expect the SPEED concept to several field.

Perhaps the most relevant limitation of the SPEED concept is the need of time varying illumination sources. For the working principle and simulations presented in here we assume pulsed light emitters with durations of the order of few ns. The short pulse duration is readily available for certain types of radiation (e.g. synchrotron X-ray radiation or Quantum Cascade Laser in the Mid Infrared region). Nevertheless, the methodology can be adapted to constant wave source with varying amplitude.

Affordable fast imaging performance at a variety of spectral regions is required in countless scientific domains and technical fields, including medicine, biology, semiconductor physics, chemistry, civil engineering, microfluidic biotechnology, transportation or security among others. Therefore, we expect the SPEED concept to impact broad areas.

# Acknowledgements

Juan Aguirre acknowledges support from the Madrid Atracción de Talento grant, project number TIC/20661.